# Field Strength Descriptions for a Confined System of SU(2) Charges

**Dennis Sivers[1]**

*Portland Physics Institute*
*Portland, OR 97239, USA*
*E-mail:* densivers@sivers.com

Applying the static Yang-Mills Maxwell equations to a simple system of SU(2) charges with spherical symmetry and confining boundary conditions provides for a demonstration of the likelihood that the confinement mechanism in non-Abelian gauge theories necessarily involves a topologically-charged domain wall consisting of a surface volume with CP-odd field strength density. The surface volume of a color-singlet system of SU(2) charges therefor describes the magnetic dual of a topological insulator. This essential topological structure is inextricably connected to the hadronic dynamics of the pion tornado. In analogy to the kink solution in the 1+1 dimensional Abelian Higgs model the classical solutions for the field equations in spherically symmetric SU(2) with this domain wall of topological charge can lead to a mass gap in the quantum system.

[1]Speaker

One of the most useful approaches to the study of non-Abelian dynamics involves the classical Bars-Witten [1,2] ansatz for the vector potential in spherically-symmetric systems of SU(2) charges,

$$gA_0^a(r,t) = A_0(r,t)\hat{r}_a$$

$$gA_i^a = A_1(r,t)\rho_{ia} + \frac{a(r,t)\sin\omega(r,t)}{r}\delta_{ia}^T + \frac{a(r,t)\cos\omega(r,t)-1}{r}\varepsilon_{ia}^T.$$

(1.1)

In this expression the tensors combine the spatial indices with adjoint SU(2) indices. Because the field equations involve the vector potential through the gauge-covariant derivative,

$$D_\mu^{ab} = \partial_\mu \delta^{ab} + g\varepsilon^{abc}A_\mu^c,$$

(1.2)

Ralston and Sivers [3-5] applied the Bars-Witten anszatz to develop a field-strength formalism that emphasizes the direct connection of spherically-symmetric SU(2) to the 1+1 dimensional Landau-Ginzburg/Abelian-Higgs model. [6] . With a change to an alternative basis for the transverse tensors,

$$e_{ia}^S(\omega) = \delta_{ia}^T \cos\omega - \varepsilon_{ia}^T \sin\omega$$

$$e_{ia}^A(\omega) = \delta_{ia}^T \sin\omega + \varepsilon_{ia}^T \cos\omega,$$

(1.3)

the expressions,

$$D_i^{ab}\hat{r}_b = \frac{a(r,t)}{r} e_{ia}^S(\omega(r,t))$$

$$-i[\hat{r}, D_i\hat{r}]_a = \frac{a(r,t)}{r} e_{ia}^A(\omega(r,t))$$

(1.4)

make it possible to write the field-strength tensor in terms of the electric and magnetic components in a gauge-covariant manner. The complete formulation of this field-strength representaton can be found in [4]. The gauge sector Lagrangian for this system is equivalent to the 1+1 dimensional Abelian Higgs model with an r-dependent metric. For the purposes of this discussion it is sufficient to consider only the static limit. The static limit of these classical expressions is of particular interest for the preliminary study of confined systems as described, for example, by E. Weinberg [7]. For the study of the static solutions to the Yang-Mills Maxwell equations we will use the convention

$$f'(r) = \frac{\partial}{\partial r} f(r).$$

(1.5)

With this convention we can express the field-strength densities,

$$E_i^a(r) = E_L(r)\rho_{ia} + E_S(r)e_{ia}^S(\omega(r)) + E_A(r)e_{ia}^A(\omega(r))$$

$$B_i^a(r) = B_L(r)\rho_{ia} + B_S(r)e_{ia}^S(\omega(r)) + B_A(r)e_{ia}^A(\omega(r)).$$

(1.6)

The components in this expression are given by

$$E_L(r) = A_0'(r) \qquad E_S(r) = \frac{a(r)A_0(r)}{r} \qquad E_A(r) = 0$$

$$B_L(r) = \frac{a^2(r)-1}{r^2} \qquad B_S(r) = -\frac{a'(r)}{r} \qquad B_A(r) = \frac{a(r)\omega'(r)}{r}$$

(1.7)

We can also define the static topological current, [4]

$$K_0(r) = (a^2(r)-1)A_1(r) - a^2(r)\omega'(r) \quad K_1(r) = -(a^2(r)-1)A_0(r) \tag{1.8}$$

The static limit of the Yang-Mills Maxwell equations is given by

$$-(r^2 E_L(r))' + 2ra(r)E_S(r) = J_0(r)$$
$$2ra(r)B_A(r) = J_1(r)$$
$$(ra(r)B_A(r))' = ra(r)j_S(r) \tag{1.9}$$

$$a(r)a''(r) - r^2(E_S^2(r) - B_A^2(r)) - \frac{a^2(r)(a^2(r)-1)}{r^2} = ra(r)j_A(r)$$

For these equations we have specified the classical currents by

$$J_0^a(r) = \frac{1}{r^2}J_0(r)\hat{r}_a$$

$$J_i^a(r) = \frac{1}{r^2}J_1(r)\rho_{ia} + j_S(r)e_{ia}^S(\omega(r)) + j_A(r)e_{ia}^A(\omega(r)), \tag{1.10}$$

The nontrivial Bianchi constraint gives the topological charge density

$$K_1'(r) = E_L(r) - [ra(r)E_S(r)]' = g^2 r^2 E_i^a B_i^a. \tag{1.11}$$

With these tools, we can quantitatively describe what a confined system of SU(2) charges would look like by directly applying the mathematical constraints given by (1.13)-(1.15) and looking for solutions with confining boundary conditions. A very simple example of this exercise starts with a solution to this system with an "interior volume" of 3-space where the vector potential is defined by the parameterization

$$A_0(r) = C_E(r)r \quad A_1(r) = C_M(r)r$$
$$a(r) = 1 \qquad \omega(r) = 0 \tag{1.12}$$

When confined to the interior volume,

$$C_E(r) = C_E \qquad r \le R_0 - \Delta$$
$$C_M(r) = C_M \qquad r \le R_0 - \Delta \tag{1.13}$$

All field parameters are constants and the static classical field-strength densities are given by

$$E_L(r) = C_E \qquad E_S(r) = C_E \qquad E_A(r) = 0$$
$$B_L(r) = 0 \qquad B_S(r) = 0 \qquad B_A(r) = C_M \tag{1.14}$$

With the classical currents specified by

$$J_0(r) = 0 \quad J_1(r) = 2rC_M \quad j_S(r) = \frac{C_M}{r} \quad j_A(r) = r(C_M^2 - C_E^2) \tag{1.15}$$

The interior volume can thus be characterized by the invariant densities,

$$E_i^a E_i^a - B_i^a B_i^a = E_L^2 - B_L^2 + 2(E_S^2 + E_A^2 - B_S^2 - B_A^2) = 3C_E^2 - 2C_M^2$$
$$E_i^a B_i^a = E_L B_L + 2(E_S B_S + E_A B_A) = 0, \tag{1.16}$$

and can therefore have a nonvanishing energy density but display vacuum quantum numbers in the SU(2) gluonic field strengths. This allows the overall quantum numbers to be determined by the quantum numbers assigned to the currents.

To apply confining boundary conditions to this simple solution, we consider an exterior volume in which all field strengths and currents vanish and a transition volume of finite thickness that contains the color gradients between the interior and exterior volumes. These gradients can then be studied using the static Yang-Mills Maxwell equations given in (1.9). Three distinct classes of solutions are considered here: type-0 given by

$$C_E^{(0)}(r) = 0 \quad C_M^{(0)}(r) = 0 \quad a^{(0)}(r) = 1 \quad \omega^{(0)}(r) = 0 \tag{1.17}$$

type-1 given by

$$C_E^{(1)}(r) = 0 \quad C_M^{(1)}(r) = 0 \quad a^{(1)}(r) = 0 \quad \omega^{(1)}(r) = 0 \tag{1.18}$$

and type-2 given by

$$C_E^{(2)}(r) = 0 \quad C_M^{(2)}(r) = 0 \quad a^{(2)}(r) = -1 \quad \omega^{(2)}(r) = \pi \tag{1.19}$$

All of these expressions are valid in the exterior volume defined by

$$r \geq R_0 + \Delta.$$

Note that type-1 solutions are not classical vacuum solutions since the exterior volume displays a radial magnetic field-strength

$$B_L^{(1)}(r) = -\frac{1}{r^2} \tag{1.20}$$

However, this exterior region can be considered a crude model for dual superconductor vacuum condensate. To interpolate between the interior volume and the exterior volume for these solutions we consider a transition region consisting of

$$r \in \left[ R_0 - \Delta, R_0 + \Delta \right]. \tag{1.21}$$

To describe this region we express the continuation in terms of the auxiliary variable

$$z = \frac{r - R_0}{\Delta} \quad z \in [-1,1]. \tag{1.22}$$

We can define two classes of interpolating functions,

$$\beta_\kappa(r) = \frac{1}{2}[1 - \tanh(\kappa z)] \quad \alpha_\kappa(r) = -\tanh(\kappa z). \tag{1.23}$$

The first of these interpolates between 1 at the boundary of the interior region and 0 at the boundary of the exterior region while the second interpolates between 1 and -1. It is also convenient to have the derivatives

$$\beta_\kappa'(r) = -\frac{1}{2}\frac{\kappa}{\cosh^2(\kappa z)} \quad \alpha_\kappa'(r) = -\frac{\kappa}{\cosh^2(\kappa z)} \tag{1.24}$$

To allow for smoothness in the continuation we can impose a series of constraints on the parameters,

$$\frac{\kappa}{R_0} \geq C_E \quad \frac{R_0}{\Delta} \geq C_E \quad \frac{\kappa}{\Delta} \geq C_E^2 \quad C_E \geq C_M. \tag{1.25}$$

The transition region for each type of solution can then be described by the two types of invariant densities. For type-0 solutions we have

$$(E_i^a E_i^a - B_i^a B_i^a)^{(0)} = C_E^2 [(\beta_\kappa + r\beta_\kappa')^2 + 2\beta_\kappa^2] - 2C_M^2 \beta_\kappa^2$$
$$(E_i^a B_i^a)^{(0)} = 0. \tag{1.26}$$

For brevity, we have suppressed the arguments in these expressions. The transition region for type-1 solutions can be described by

$$(E_i^a E_i^a - B_i^a B_i^a)^{(1)} = \frac{1}{r^4}[r^4 C_E^2 (\beta_\kappa + r\beta_\kappa')^2 + 2r^4 (C_E^2 - C_M^2)\beta_\kappa^4 - 2r^2 \beta_\kappa'^2 - (\beta_\kappa^2 - 1)^2]$$

$$(E_i^a B_i^a)^{(1)} = \frac{C_E}{r^2}[(\beta_\kappa^2 - 1)(\beta_\kappa + r\beta_\kappa') - 2r\beta_\kappa^2 \beta_\kappa']$$

$$\tag{1.27}$$

The transition region for type-2 solutions is then given by

$$(E_i^a E_i^a - B_i^a B_i^a)^{(2)} = C_E^2 [(\beta_\kappa + r\beta_\kappa')^2 + 2\alpha_\kappa^2 \beta_\kappa^2] + \frac{1}{r^4}[(\alpha_\kappa^2 - 1)^2 + 2r^2 (\alpha_\kappa'^2 + \alpha_\kappa^2 (rC_M \beta_\kappa - \pi\beta_\kappa')^2]$$

$$(E_i^a B_i^a)^{(2)} = \frac{C_E}{r^2}[(\alpha_\kappa^2 - 1)(\beta_\kappa + r\beta_\kappa') - 2r\alpha_\kappa \alpha_\kappa' \beta_\kappa]$$

$$\tag{1.28}$$

The color gradients in the intermediate region described here, in turn, induce color currents in the region creating a color riparian zone that will be discussed in a larger paper. The absence of topological charge in the type-0 solutions implies that these solutions are unstable with respect to small fluctuations that can allow color to migrate into the exterior region. For type-1 solutions, the CP-odd quantum numbers of the transition region provide a topological cushion separating the color-charged interior volume from the color-charge repelling exterior volume. For the type-2 solutions the domain wall of topological charge separates two regions with "vacuum" quantum numbers, in analogy to the kink solution of the 1+1 dimensional Abelian-Higgs model

$$a^{(2)}(r) = 1 \quad \omega^{(2)}(r) = 0 \quad r \leq R_0 - \Delta$$
$$a^{(2)}(r) = 1 \quad \omega^{(2)}(r) = \pi \quad r \geq R_0 + \Delta \tag{1.29}$$

Both type-1 and type-2 solutions can provide interesting starting points for further study.